\documentstyle[12pt]{article}
\begin{document}
\centerline{\Large\bf On the Meaning of Various Mass Definitions}
\centerline{\Large\bf for Asymptotically Flat Spacetimes}
\vskip .7in
\centerline{Dan N. Vollick}
\vskip .2in
\centerline{Irving K. Barber Faculty of Science}
\centerline{University of British Columbia Okanagan}
\centerline{3333 University Way}
\centerline{Kelowna, B.C.}
\centerline{Canada}
\centerline{V1V 1V7}
\vskip 0.5in
\centerline{\bf\large Abstract}
\vskip 0.5in
\noindent
The mass contained in an arbitrary spacetime in general relativity is not well defined. However, for asymptotically flat spacetimes various definitions of mass have been proposed.  In this paper I consider eight masses and show that some of them correspond to the active gravitational mass while the others correspond to the inertial mass. For example, the ADM mass corresponds to the inertial mass while the M$\o$ller mass corresponds to the active gravitational mass.  In general the inertial and active gravitational masses are not equal. If the spacetime is vacuum at large $r$ the Einstein equations force the inertial and active gravitational masses to be the same. The Einstein equations also force the masses to be the same if any matter that extends out to large $r$ satisfies the weak, strong or dominant energy condition.
I also examine the contributions of the inertial and active gravitational masses to the gravitational redshift, the deflection of light, the Shapiro time delay, the precession of perihelia and to the motion of test bodies in the spacetime.
\newpage
\section{Introduction}
In general relativity he mass contained in an arbitrary spacetime is not well defined. However, for asymptotically flat spacetimes various definitions of mass have been proposed. In this paper I consider eight masses: The active gravitational mass, the Einstein mass, the Landau-Lifshitz mass, the ADM mass, the M$\o$ller mass, the Tolman mass, the Komar mass and the $\rho$-mass in asymptotically flat spacetimes. The metric is taken to be
\begin{equation}
ds^2\simeq-\left[1-\frac{2GM_1}{r}\right]dt^2+\left[1+\frac{2GM_2}{r}\right]dr^2+r^2d\Omega^2
\label{1}
\end{equation}
at large $r$, where $M_1$ and $M_2$ are constants.  Spacetimes with $M_1\neq M_2$ occur in string theory \cite{Gar1} and in Brans-Dicke theory \cite {Brans1}.
Some of the above masses correspond to $M_1$, while others correspond to $M_2$. The physical significance of $M_1$ is well known from the Newtonian limit. It is the active gravitational mass of the system. But, what is the physical significance of $M_2$? If the spacetime is vacuum at large distances we must have $M_1=M_2$, but this won't be the case if the matter distribution extends out to infinity. In this paper I examine the physical meaning of $M_2$ and argue that it corresponds to the inertial mass of the system. This identification is made by considering the mass, associated with the energy-momentum tensor, that flows in from infinity in a spacetime that evolves from a Minkowski spacetime into the spacetime described by the metric (\ref{1}). The mass that flows in is $M_2$ and since the energy-momentum tensor contains the inertial mass, not the gravitational mass $M_2$ is the inertial mass. Some of the masses, therefore, correspond to the inertial mass, while the others correspond to the active gravitational mass. It is interesting to note that if $M_1\neq M_2$ the inertial and active gravitational masses are not equal. This does not imply a violation of the weak equivalence principle which requires the equality of inertial and passive gravitational masses.
  
The paper is organized as follows. In section 2 the above mass definitions are discussed in detail. In section 3 the masses for the spacetime with metric (\ref{1}) are computed and the physical significance of the parameters $M_1$ and $M_2$ are investigated. I also
examine the contributions of $M_1$ and $M_2$ to the gravitational redshift, the deflection of light, the Shapiro time delay and the precession of perihelia. The results of the paper are summarized in section 4.
\section{Discussion of the Various Mass Definitions}
In this section I will consider eight definitions of the mass for asymptotically flat spacetimes.
\newline
\noindent \textbf{The Active Gravitational Mass.} In weak gravitational fields, where $g_{\mu\nu}=\eta_{\mu\nu}+h_{\mu\nu}$ and $|h_{\mu\nu}|<<1$, the geodesic equation is given by
\begin{equation}
\frac{d^2\vec{r}}{dt^2}\simeq\frac{1}{2}\vec{\nabla}h_{tt}.
\end{equation}
Comparing this to the Newtonian equation gives
\begin{equation}
h_{tt}=\frac{2GM_G}{r}\;,
\label{MG}
\end{equation}
where $M_G$ is the active gravitational mass. In asymptotically flat spacetimes
\begin{equation}
g_{tt}\rightarrow-\left(1-\frac{2GM_G}{r}\right)\;\;\;\;\;\;\; as \;\;\;\;\;\;\;\; r\rightarrow\infty
\end{equation}
allowing $M_G$ to be easily identified.
\newline
\noindent\textbf{The $\rho$-mass.} Consider a spherically symmetric asymptotically flat static spacetime. The Einstein equation $G_{tt}=-8\pi G T_{tt}$ gives
\begin{equation}
\frac{1}{r^2}\frac{d}{dr}\left[r\left(1-g_{rr}^{-1}\right)\right]=8\pi G\rho\;.
\end{equation}
the solution to this equation is
\begin{equation}
g_{rr}(r)=\left[1-\frac{2Gm(r)}{r}\right]^{-1}
\end{equation}
where
\begin{equation}
m(r)=4\pi\int_0^r\rho(r')(r')^2dr'\;.
\end{equation}
The $\rho$-mass is defined to be
\begin{equation}
M_{\rho}=4\pi\int_0^{\infty}\rho(r)r^2dr\;.
\end{equation}
Thus, for large $r$
\begin{equation}
g_{rr}\simeq 1+\frac{2GM_{\rho}}{r}
\label{MR}
\end{equation}
allowing $M_{\rho}$ to be easily identified. Note that the spacetime must be spherically symmetric for $M_{\rho}$ to be defined.
\newline
\noindent \textbf{The Einstein, Landau-Lifshitz and ADM Masses.} Einstein showed that the conservation laws $\nabla_{\nu}T^{\;\;\nu}_{\mu}=0$ can be written in the form
\begin{equation}
\frac{\partial}{\partial x^{\nu}}\left[\sqrt{-g}\left(T^{\;\;\nu}_{\mu}+t^{\;\;\nu}_{\mu}\right)\right]=0\;,
\end{equation}
where $t^{\;\;\nu}_{\mu}$ is the Einstein or canonical energy-momentum pseudotensor and is given by
\begin{equation}
t^{\;\;\nu}_{\mu}=\frac{1}{2\kappa\sqrt{-g}}\left[\frac{\partial (\sqrt{-g}L)}{\partial g^{\alpha\beta}_{\;\;\;\;\;,\nu}}g^{\alpha\beta}_{\;\;\;\;\;,\mu}-
\delta^{\;\;\nu}_{\mu}\sqrt{-g}L\right]\;,
\end{equation}
where
\begin{equation}
L=g^{\mu\nu}\left(\Gamma^{\alpha}_{\mu\nu}\Gamma^{\beta}_{\alpha\beta}-\Gamma^{\alpha}_{\mu\beta}\Gamma^{\beta}_{\nu\alpha}\right)\;.
\end{equation}
If the spacetime is foliated into spacelike hypersurfaces the quantity
\begin{equation}
P_{\mu}=\int \left[\sqrt{-g}\left(T^{\;\;t}_{\mu}+t^{\;\;t}_{\mu}\right)\right]d^3x
\label{mom}
\end{equation}
is the same on each hypersurface, if the system is closed.  It is not possible to think of $t^{\;\;\nu}_{\mu}$ as the energy-momentum tensor of the gravitational field since it is not a tensor. For example, Bauer \cite{Bau} has shown that $t^{\;\;\nu}_{\mu}$ does not vanish in flat spacetime if spherical coordinates are used and that the total energy of flat spacetime in spherical coordinates diverges.

Freud \cite{Fre} has shown that
\begin{equation}
\sqrt{-g}\left(T^{\;\;\nu}_{\mu}+t^{\;\;\nu}_{\mu}\right)=\frac{\partial (\sqrt{-g}U_{\mu}^{\;\;\nu\alpha})}{\partial x^{\alpha}}
\label{potential1}
\end{equation}
and M$\o$ller \cite{Mol1} has shown that the $U_{\mu}^{\;\;\nu\alpha}$ can be written as
\begin{equation}
\sqrt{-g}U_{\mu}^{\;\;\nu\alpha}=\frac{g_{\mu\beta}}{2\kappa \sqrt{-g}}\left[(-g)\left(g^{\nu\beta}g^{\alpha\lambda}-g^{\alpha\beta}g^{\nu\lambda}\right)\right]_{,\lambda}\;.
\label{Mo}
\end{equation}
The quantity $P_{\mu}$ can now be written as a surface integral at infinity
\begin{equation}
P_{\mu}=\int \sqrt{-g}U_{\mu}^{\;\; tk}dS_{k}\;.
\end{equation}
If the spacetime is asymptotically flat and coordinates are chosen so that the $g_{\mu\nu}=\eta_{\mu\nu}+h_{\mu\nu}$ at large distances with $h_{\mu\nu}$ dropping off as $1/r$ or faster the integral will give a finite result. It is also easy to see that $P_{\mu}$ is left unchanged by arbitrary coordinate transformations that approach the identity at infinity. $P_{\mu}$ also transforms as a four vector under Lorentz transformations. These properties of $P_{\mu}$ make it a candidate for the total energy and momentum of the system.
\newline
Landau and Lifshitz \cite {LL1} have shown that
\begin{equation}
(-g)\left(T^{\mu\nu}+t^{\mu\nu}\right)=\frac{\partial h^{\mu\nu\alpha}}{\partial x^{\alpha}}\;,
\end{equation}
where
\begin{equation}
h^{\mu\nu\alpha}=\frac{1}{16\pi G}\left[(-g)\left(g^{\mu\nu}g^{\alpha\beta}-g^{\mu\alpha}g^{\nu\beta}\right)\right]_{,\beta}\;.
\label{LLh}
\end{equation}
The antisymmetry of $h^{\mu\nu\alpha}$ in $\nu$ and $\alpha$ implies that
\begin{equation}
\frac{\partial}{\partial x^{\nu}}\left[(-g)\left(T^{\mu\nu}+t^{\mu\nu}\right)\right]=0
\end{equation}
which in turn implies that
\begin{equation}
P^{\mu}=\int(-g)\left(T^{\mu\nu}+t^{\mu\nu}\right)dS_{\nu}
\end{equation}
is a conserved quantity. 
\newline
Arnowitt, Deser and Misner \cite{ADM} constructed the Hamiltonian for general relativity and identified it with the total energy. In asymptotically flat coordinates the ADM mass is given by
\begin{equation}
M_{ADM}=\frac{1}{16\pi G}\int\left(g_{ik,k}-g_{kk,i}\right)dS^i\;.
\label{MADM}
\end{equation}
This mass also follows from the energy-momentum pseudo-tensor discussed by Weinberg \cite{Wei1}. In this approach the metric is written as
\begin{equation}
g_{\mu\nu}=\eta_{\mu\nu}+h_{\mu\nu}\;,
\end{equation}
where $h_{\mu\nu}$ vanishes at infinity, but is not necessarily small everywhere. The Einstein equations are written as
\begin{equation}
G_{\mu\nu}^{(1)}=-8\pi G\left(T_{\mu\nu}+t_{\mu\nu}\right)\;,
\end{equation}
where $G_{\mu\nu}^{(1)}$ is the part of the Einstein tensor that is linear in $h_{\mu\nu}$ and
\begin{equation}
t_{\mu\nu}=\frac{1}{8\pi G}\left[G_{\mu\nu}-G_{\mu\nu}^{(1)}\right]\;.
\end{equation}
The mass associated with $t_{\mu\nu}$ is identical to the ADM mass. The pseudo-tensor $t^{\mu\nu}$ can be written as
\begin{equation}
t^{\mu\nu}=\partial_{\alpha}\tilde{Q}^{\alpha\nu\alpha}\;,
\end{equation}
where I have modified Weinberg's notation ($\tilde{Q}^{\alpha\nu\alpha}$ is given below in equation (\ref{potential2})).

Consider an asymptotically flat spacetime with $g_{\mu\nu}=\eta_{\mu\nu}+h_{\mu\nu}$, where $h_{\mu\nu}$ vanishes at infinity. In the asymptotic region
\begin{equation}
\sqrt{-g}U_{\mu}^{\;\;\nu\alpha}=h_{\mu}^{\;\;\nu\alpha}=\tilde{Q}_{\mu}^{\;\;\nu\alpha}=\frac{1}{2\kappa}\left\{\delta^{\nu}_{\;\;\mu}\left[\partial^{\alpha}h
-\partial_{\beta}h^{\beta\alpha}\right]-\delta^{\alpha}_{\;\;\mu}\left[\partial^{\nu}h-\partial_{\beta}h^{\beta\nu}\right]+\partial^{\nu}h^{\alpha}_{\;\;\mu}-
\partial^{\alpha}h^{\nu}_{\;\;\mu}\right\}
\label{potential2}
\end{equation}
implying that the Einstein, Landau-Lifshitz and ADM masses are the same.
\newline
\noindent \textbf{The M$\bf{\o}$ller and Tolman Masses.} M$\o$ller \cite{Mol1} noted that $P_{\mu}$ is not invariant under coordinate transformations on the $t=$constant hypersurfaces because $\sqrt{-g}\left(T^{\;\;t}_{\mu}+t^{\;\;t}_{\mu}\right)$ is not a 4-vector density. To correct this he used the freedom to modify the energy-momentum tensor by adding $S^{\;\;\nu}_{\mu}$ to
$\Theta^{\;\;\nu}_{\mu}=\sqrt{-g}\left(T^{\;\;\nu}_{\mu}+t^{\;\;\nu}_{\mu}\right)$, where $\partial_{\nu}S^{\;\;\nu}_{\mu}=0$. M$\o$ller's energy-momentum pseudo-tensor is given by
\begin{equation}
J^{\;\;\nu}_{\mu}=\Theta^{\;\;\nu}_{\mu}+S^{\;\;\nu}_{\mu}\;.
\end{equation}
where $S^{\;\;\nu}_{\mu}$ was chosen so that $J^{\;\;t}_{\mu}=\Theta^{\;\;t}_{\mu}+S^{\;\;t}_{\mu}$ is a 4-vector density. In addition M$\o$ller showed that
\begin{equation}
J^{\;\;\nu}_{\mu}=\frac{\partial\chi^{\;\;\nu\alpha}_{\mu}}{\partial x^{\alpha}}\;,
\end{equation}
where
\begin{equation}
\chi^{\;\;\nu\alpha}_{\mu}=\frac{\sqrt{-g}}{8\pi G}\left[\frac{\partial g_{\mu\beta}}{\partial x^{\lambda}}-\frac{\partial g_{\mu\lambda}}{\partial x^{\beta}}\right]g^{\nu\lambda}
g^{\alpha\beta}\;.
\end{equation}
The energy and momentum that follows from M$\o$ller's energy-momentum pseudo-tensor are given by
\begin{equation}
P_{\mu}=\int \chi_{\mu}^{\;\; tk}dS_{k}\;.
\label{MM}
\end{equation}
M$\o$ller has shown that the total mass of a static spacetime that follows from $J^{\;\;t}_{\mu}$ can be written as
\begin{equation}
M_M=\int\left(T^{\;\;k}_{k}-T^{\;\;t}_{t}\right)\sqrt{-g}d^3x\;.
\end{equation}
This is the same expression that was derived by Tolman \cite{Tol1}. The Tolman and M$\o$ller masses are, therefore, the same in static spacetimes (see \cite{Flo1}).

In the asymptotic region $\chi^{\;\;\nu\alpha}_{\mu}$ is given by
\begin{equation}
\chi^{\;\;\nu\alpha}_{\mu}=\frac{1}{\kappa}\left[\partial^{\nu}h_{\mu}^{\;\;\alpha}-\partial^{\alpha}h_{\mu}^{\;\;\nu}\right]\;.
\end{equation}
\newline
\noindent \textbf{The Komar Mass.} In an asymptotically flat spacetime with a timelike Killing vector $\xi^{\mu}$ the quantity
\begin{equation}
J^{\mu}=R^{\mu\nu}\xi_{\nu}
\end{equation}
satisfies
\begin{equation}
\nabla_{\mu}J^{\mu}=0\;.
\end{equation}
This continuity equations allows a conserved energy
\begin{equation}
M_K=\frac{1}{4\pi G}\int_{\Sigma}\sqrt{\gamma}n_{\mu}J^{\mu}d^3x\;,
\end{equation}
to be defined, where $n^{\mu}$ is the normal to the hypersurface $\Sigma$ and $M_K$ is the Komar mass \cite{Kom1,Car1}. The volume integral can be converted into a surface integral giving
\begin{equation}
M_K=\frac{1}{4\pi G}\int_{\partial\Sigma}\sqrt{\gamma^{(2)}}n_{\mu}\sigma_{\nu}\nabla^{\mu}\xi^{\nu}d^2x\;,
\label{MK}
\end{equation}
where $\sigma^{\mu}$ is the normal to $\partial\Sigma$. In the asymptotic region $n_{\mu}\sigma_{\nu}\nabla^{\mu}\xi^{\nu}$ is given by
\begin{equation}
n_{\mu}\sigma_{\nu}\nabla^{\mu}\xi^{\nu}=-\frac{x^k}{2r}\partial_kh_{tt}\;.
\end{equation}
\section{Comparison of the Masses in Asymptotically Flat Spacetimes}
Consider a static asymptotically flat spacetime with a metric given by
\begin{equation}
ds^2\simeq-\left[1-\frac{2GM_1}{r}\right]dt^2+\left[1+\frac{2GM_2}{r}\right]dr^2+r^2d\Omega^2\;.
\label{metric}
\end{equation}
at large $r$, where $M_1$ and $M_2$ are constants. Note that the spacetime does not have to be spherically symmetric, but only have the above asymptotic form.

The energy-momentum tensor can be found from the Einstein equations
\begin{equation}
T^{\mu}_{\;\;\;\nu}=-\frac{1}{8\pi G}G^{\mu}_{\;\;\;\nu}.
\end{equation}
The non-zero components, at large $r$, are given by
\begin{equation}
T^{t}_{\;\;\;t}\simeq -\frac{M_2^2}{4\pi r^4}\;,
\end{equation}
\begin{equation}
T^{r}_{\;\;\;r}\simeq \frac{(M_1-M_2)}{4\pi r^3}\;,
\end{equation}
and
\begin{equation}
T^{\theta}_{\;\;\;\theta}=T^{\phi}_{\;\;\;\phi}\simeq \frac{(M_2-M_1)}{8\pi r^3}\;.
\end{equation}
The expression for $T^{t}_{\;\;\;t}$ is sensitive to higher order terms in $g_{rr}$ (i.e. terms that fall off as $1/r^2$ or faster). For example if $g_{rr}=(1-2GM_2/r)^{-1}$ then $T^{t}_{\;\;\;t}=0$. It can be shown that $T^{t}_{\;\;\;t}$ will always fall off as $1/r^4$ or faster.
The leading order terms in $T^{r}_{\;\;\;r}, T^{\theta}_{\;\;\;\theta}$ and $T^{\phi}_{\;\;\;\phi}$ are insensitive to higher order terms in $g_{tt}$ and $g_{rr}$.
It is interesting to note that this energy-momentum tensor violates the weak, strong and dominant energy condition if $M_1\neq M_2$. This implies that any field, such as the electromagnetic field or the Klein-Gordon scalar field, that satisfies any of the energy conditions, cannot produce a spacetime with $M_1\neq M_2$. Spacetimes with $M_1\neq M_2$ do occur in string theory. For example, the string metric for a charged dilaton black hole is given by (G=1)\cite{Gar1}
\begin{equation}
ds_{string}^2=-\frac{(1-2Me^{\phi_0}/r)}{(1-Q^2e^{3\phi_0}/Mr)}dt^2+\frac{dr^2}{(1-2Me^{\phi_0}/r)(1-Q^2e^{3\phi_0}/Mr)}+r^2d\Omega^2.
\end{equation}
The values of $M_1$ and $M_2$ are given by
\begin{equation}
M_1=Me^{\phi_0}-\frac{Q^2e^{3\phi_0}}{2M}\;\;\;\;\;\;\;\;\;\; and \;\;\;\;\;\;\;\;\;\;
M_2=Me^{\phi_0}+\frac{Q^2e^{3\phi_0}}{2M}.
\end{equation}
If $Q\neq 0$ we see that $M_1\neq M_2$. In fact,
for an extremal black hole ($Q^2=2M^2e^{-2\phi_0}$) $M_1$ vanishes and $M_2=2Me^{\phi_0}$. Spacetimes with $M_1\neq M_2$ also occur in Brans-Dicke theory \cite{Brans1}.

The masses discussed in the previous section can be found from (\ref{MG}), (\ref{MR}), (\ref{MADM}), (\ref{MM}), (\ref{MK}) and
\begin{equation}
h_{tt}=\frac{2GM_1}{r}\;\;\;\;\;\;\;\;\;\; and \;\;\;\;\;\;\;\;\;\; h_{ij}=\frac{2G M_2}{r^3}x_ix_j\;.
\end{equation}
They are given by
\begin{equation}
M_G=M_M=M_T=M_K=M_1 \;\;\;\;\;\;\;\;\;\;\; and \;\;\;\;\;\;\;\;\;\;\;\;\; M_{\rho}=M_E=M_{LL}=M_{ADM}=M_2\;,
\end{equation}
where $M_G$ is the active gravitational mass, $M_M$ is the M$\o$ller mass, $M_T$ is the Tolamn mass, $M_K$ is the Komar mass, $M_{\rho}$ is the $\rho$-mass, $M_E$ is the Einstein mass, $M_{LL}$ is the Landau-Lifshitz mass and $M_{ADM}$ is the ADM mass. Recall that the $\rho$-mass is defined only in spherically symmetric spacetimes.

From $M_G=M_1$ it is clear that $M_1$ is the active gravitational mass of the system. What then does $M_2$ represent? To answer this question
consider a time dependent asymptotically flat spacetime with a metric given by
\begin{equation}
ds^2\simeq-\left[1-\frac{2GM_1(t)}{r}\right]dt^2+\left[1+\frac{2 GM_2(t)}{r}\right]dr^2+r^2d\Omega^2
\label{metric2}
\end{equation}
at large $r$ where
\begin{equation}
\lim_{t\rightarrow-\infty}M_{1,2}(t)=0
\end{equation}
and
\begin{equation}
\lim_{t\rightarrow\infty}M_{1,2}(t)=M_{1,2}.
\end{equation}
Alternatively $M_1(t)$ and $M_2(t)$ can be taken to be
\begin{equation}
M_{1,2}(t)= \left\{\begin{array}{cc}
                           0\;\;\;\;\;\;\;\;\;\; t\leq t_1& \\
                           \tilde{M}_{1,2}(t)\;\;\;\;\; t_1<t<t_2 & \\
                           M_{1,2}\;\;\;\;\;\;\;\;\;\; t\geq t_2
                          \end{array}\hspace {-0.12in}\right\}\;.
\label{mass}
\end{equation}
where $M_{1,2}(t)\in C^n$, with $n\geq 2$. This spacetime starts as a Minkowski spacetime and evolves into the spacetime discussed earlier which has a metric given by (\ref{metric}) at large $r$.

The non-zero components of the energy-momentum tensor, at large $r$, are given by
\begin{equation}
T^{t}_{\;\;\;t}\simeq -\frac{M_2(t)^4}{4\pi r^4}\;,
\end{equation}
\begin{equation}
T^{r}_{\;\;\;r}\simeq \frac{\left[M_1(t)-M_2(t)\right]}{4\pi r^3}
\end{equation}
\begin{equation}
T^{\theta}_{\;\;\;\theta}=T^{\phi}_{\;\;\;\phi}\simeq -\frac{1}{8\pi r}\frac{d^2M_2(t)}{dt^2}\;,
\end{equation}
and
\begin{equation}
T^{t}_{\;\;\;r}\simeq -\frac{1}{4\pi  r^2}\frac{dM_2(t)}{dt}\;,
\end{equation}
Once again, $T^{t}_{\;\;\;t}$ is sensitive to higher order terms in $g_{rr}$ while the other components are not sensitive to higher order terms in $g_{tt}$ or $g_{rr}$. Note that $T^{t}_{\;\;\;r}\neq 0$. This implies that mass will flow in from infinity.

Consider a sphere of fixed coordinate radius $R$ with $R\rightarrow\infty$. Observers at rest on this sphere will measure an energy flux given by $T^{t}_{\;\;\;r}$ and the total rate at which energy flows in through the sphere is given by
\begin{equation}
\frac{dE}{dt}=\frac{dM_2(t)}{dt}\;.
\end{equation}
The total energy that flows in through the sphere is, therefore, given by
\begin{equation}
E=M_2\;.
\end{equation}
  
The energy-momentum tensor $T^{\mu\nu}$ contains the inertial mass, not the gravitational mass. To see this consider charged dust coupled to the electromagnetic field.
The energy-momentum tensor for the system is given by
\begin{equation}
T^{\mu\nu}=\rho_m U^{\mu}U^{\nu}+F^{\mu}_{\;\;\alpha}F^{\nu\alpha}-\frac{1}{4}g^{\mu\nu}F_{\alpha\beta}F^{\alpha\beta},
\end{equation}
where $\rho_m$ is the mass density of the dust, $U^{\mu}$ is its four-velocity and $F^{\mu\nu}$ is the electromagnetic field tensor. The equations of motion that follow from $\nabla_{\mu}T^{\mu\nu}=0$ and the field equations
\begin{equation}
\nabla_{\mu}F^{\mu\nu}=-\rho_cU^{\nu}\;\;\;\;\;\;\;\;\;\;\; and \;\;\;\;\;\;\;\;\;\;\;\; \nabla_{[\mu}F_{\alpha\beta]}=0
\end{equation}
are
\begin{equation}
\rho_mU^{\nu}\nabla_{\nu}U^{\mu}=\rho_cF^{\mu}_{\;\;\;\nu}U^{\nu},
\end{equation}
where $\rho_c$ is the charge density of the dust and $[\cdot\cdot\cdot]$ denotes antisymmetrization.
The mass density $\rho_m$ is obviously the inertial mass density. Therefore, an amount of inertial mass equal to $M_2$ flowed into the system. The inertial mass may, of course, contribute to the active gravitational mass, but there  is no a priori reason that they must be equal.

There are two independent conservation laws:
\begin{equation}
\partial_{\nu} J^{\;\;\nu}_{\mu}=0
\label{conserv2}
\end{equation}
and
\begin{equation}
\frac{\partial}{\partial x^{\nu}}\left[\sqrt{-g}\left(T^{\;\;\nu}_{\mu}+t^{\;\;\nu}_{\mu}\right)\right]=0
\label{conserv1}
\end{equation}
where $ J^{\;\;\nu}_{\mu}$ is the  M$\o$ller pseudotensor and  $t^{\;\;\nu}_{\mu}$ is the Einstein pseudotensor. The Landau-Lifshitz or Weinberg pseudotensor could be used instead of the Einstein pseudotensor since they all give the same mass. Since $M_M=M_G$, I postulate that 
({\ref{conserv2}) corresponds to the conservation of the active gravitational mass. Since $M_E$ corresponds to the inertial mass that flowed into the system, I postulate that (\ref{conserv1}) corresponds to the conservation of inertial mass. I therefore identify $M_2$ as the inertial mass of the system. Note that the inertial and active gravitational masses do not have to be the same, but both are individually conserved. This does not imply a violation of the weak equivalence principle which requires the equality of inertial and passive gravitational masses. The Einstein equations force the inertial and active gravitational masses to be the same if the matter that extends out to large $r$ satisfies the weak, strong or dominant energy condition.  If the spacetime has a metric of the form (\ref{metric}) and is vacuum at large $r$ the Einstein equations also force the inertial and active gravitational masses to be the same. Bonner \cite{Bon1} and Rosen and Cooperstock \cite{Ros1} discuss the equality of inertial, active and passive gravitational masses for bodies composed of ideal fluids. Bonner claims that the active gravitational mass does not equal the passive gravitational mass (which equals the inertial mass). Rosen and Cooperstock claim that, if the gravitational self energy is included, all three masses are the same. In both articles it is assumed that $M_1=M_2$.
For the remainder of this paper (except in the conclusion) I will denote $M_1$ by $M_G$ and $M_2$ by $M_{ADM}$. I have chosen to denote $M_2$ by $M_{ADM}$ instead of $M_I$ (the inertial mass) because the ADM mass is commonly used in the literature.
  
Ohanian \cite{Ohan1,Ohan2,Ohan3} comes to the same conclusion
(i.e. $M_2$ equals the inertial mass) by identifying the volume integral over the canonical energy-momentum pseudotensor of the matter and gravitational field as the inertial mass.
There can be, however, a problem with using the canonical energy-momentum pseudotensor. Consider, for example, the electromagnetic field in flat spacetime. The canonical energy-momentum pseudotensor $\Theta_{\mu}^{\;\;\nu}$ is related to the standard energy-momentum tensor $T_{\mu}^{\;\;\nu}$ by
\begin{equation}
\Theta_{\mu}^{\;\;\nu}=T_{\mu}^{\;\;\nu}-\partial_{\alpha}\left(F^{\alpha\nu}A_{\mu}\right)\;.
\label{can}
\end{equation}
Note that $\Theta_{\mu\nu}$ is neither symmetric nor gauge invariant. Ohanian argues that the total energy derived from $\Theta_{\mu}^{\;\;\nu}$ is the same as the total energy derived from $T_{\mu}^{\;\;\nu}$. This can only be true if the last term in (\ref{can}) does not contribute to the energy. The contribution of this term to the energy is given by
\begin{equation}
\int\partial_{k}\left(F^{tk}A_{t}\right)d^3x=\int\left(F^{tk}A_{t}\right)dS_k
=-\int\phi\vec{E}\cdot d\vec{a}\;.
\label{Ohan}
\end{equation}
This term will vanish if the fields drop off sufficiently rapidly at infinity, which Ohanion assumed. This term, however, is not necessarily zero and is not gauge invariant.
Consider, for example the two sets of fields
\begin{equation}
\vec{E}=\frac{q}{r^2}\hat{r},\;\;\;\;\;\;\;\;\;\;\phi=\frac{q}{r} \;\;\;\;\;\;\;\;\;\;\vec{A}=0
\end{equation}
and
\begin{equation}
\vec{E}=\frac{q}{r^2}\hat{r},\;\;\;\;\;\;\;\;\;\;\phi=\frac{q}{r}-\chi(t) \;\;\;\;\;\;\;\;\;\;\vec{A}=0\;
\end{equation}
which are related by a gauge transformation. The integral in (\ref{Ohan}) vanishes for the first set of fields but is given by
\begin{equation}
\int\partial_{k}\left(F^{tk}A_{t}\right)d^3x=q\chi(t)
\end{equation}
for the second set of fields. This term is neither zero nor gauge invariant. Note that this term is not time independent, which implies that the four-momentum
\begin{equation}
P_{\mu}=\int\Theta_{\mu}^{\;\;\; t}d^3x
\end{equation}
is not conserved. To see why this is the case consider
\begin{equation}
\frac{dP_{\mu}}{dt}=\int\frac{\partial\Theta_{\mu}^{\;\;\; t}}{\partial t}d^3x=-\int\partial_k\Theta_{\mu}^{\;\;\; k}d^3x=-\int\Theta_{\mu}^{\;\;\; k}dS_k\neq 0.
\end{equation}
The canonical energy-momentum tensor, therefore, cannot be used in general.
  
Now consider the motion of a test mass at large $r$. In isotropic coordinates
\begin{equation}
ds^2\simeq-\left[1-\frac{2GM_G}{\rho}\right]dt^2+\left[1+\frac{2GM_{ADM}}{\rho}\right]\left[dx^2+dy^2+dz^2\right]
\label{iso}
\end{equation}
at large $\rho$, where $r=\rho(1+GM_{ADM}/2\rho)^2$.
The equations of motion are
\begin{equation}
\left[1+\frac{2GM_{ADM}}{c^2\rho}\right]\frac{d^2\vec{\rho}}{d\tau^2}=-\left[\frac{GM_G}{\rho^2}\hat{\rho}\right]\dot{t}^2-\frac{GM_{ADM}}{c^2\rho^2}\left[v^2\hat{\rho}-2(\hat{\rho}\cdot\vec{v})\vec{v}\right]
\end{equation}
and
\begin{equation}
\left[1-\frac{2GM_G}{c^2\rho}\right]c^2\dot{t}^2-\left[1+\frac{2GM_{ADM}}{c^2\rho}\right]v^2=1,
\end{equation}
where $\dot{t}=dt/d\tau$, $\vec{v}=d\vec{\rho}/d\tau$ and $\vec{\rho}=(x,y,z)$. Note that higher order terms in the metric (\ref{iso}) will produce additional terms in the above equations of motion. The gravitational mass gives the expected contribution, but the ADM mass only produces corrections to the Newtonian force. If $M_G=0$ a particle at rest will not experience a ``force". This is hidden in Schwarzschild spacetimes because the ADM mass has the same value as the gravitational mass.
  
It is also interesting to examine the contributions of $M_G$ and $M_{ADM}$ to the gravitational redshift, the deflection of light, the Shapiro time delay and the precession of perihelia.
Consider two observers, one at position $x_1^{\mu}$ and the second at $x_2^{\mu}$. If the first observer sends a light signal with frequency $\nu_1$ the second observer will measure its frequency to be \cite{Wei1}
\begin{equation}
\nu_2=\sqrt{\frac{g_{tt}(x_2)}{g_{tt}(x_1)}}\nu_1\;.
\end{equation}
To lowest order
\begin{equation}
\frac{\Delta\nu}{\nu}\simeq \frac{GM_G}{r_1}-\frac{GM_G}{r_2},
\end{equation}
which is the result obtained from the equivalence principle or by considering a photon moving in a Newtonian gravitational field. It is interesting to consider the gravitational redshift in isotropic coordinates. Transforming (\ref{metric}) to isotropic coordinates gives
\begin{equation}
g_{tt}^{(iso)}=-\left(1-\frac{2GM_G}{\rho}+\frac{2M_GM_{ADM}}{\rho^2}+\cdot\cdot\cdot\right).
\end{equation}
In isotropic coordinates the gravitational redshift, to lowest order, depends only on $M_G$, but higher order corrections depend on $M_{ADM}$. This results from the fact that the transformation between $r$ and $\rho$ involves $M_{ADM}$. One could also consider the transformation $r=\frac{M_G}{M_{ADM}}\bar{r}$ giving $\bar{g}_{tt}=-(1-2GM_{ADM}/{\bar{r}})$. In this coordinate system the gravitational redshift depends on $M_{ADM}$ not $M_G$. However, in this coordinate system the metric does not go to the Minkowski metric, in spherical coordinates, at large $\bar{r}$. If the radial coordinate transformation is restricted to transformations that satisfy $\bar{r}\rightarrow r$ as $r\rightarrow\infty$ the lowest order term will depend only on $M_G$.
  
To examine light deflection it is convenient to use the Eddington-Robertson expansion
\begin{equation}
ds^2=-\left(1-\frac{2M}{r}\right)dt^2+\left(1+\frac{2\gamma GM}{r}\right)dr^2+r^2d\Omega^2.
\end{equation}
To match the metric (\ref{metric}) with $M_1=M_G$ and $M_2=M_{ADM}$ requires that
\begin{equation}
M=M_G,\;\;\;\;\;\;\;\;\; and \;\;\;\;\;\;\;\;\; M_{ADM}=\gamma M\;.
\end{equation}
In terms of the Eddington-Robertson parameters the bending of light, to lowest order, is given by
\begin{equation}
\Delta\theta=\frac{4GM}{b}\left(\frac{1+\gamma}{2}\right),
\end{equation}
where $b$ is the impact parameter. In terms of $M_G$ and $M_E$ this becomes
\begin{equation}
\Delta\theta=\frac{4G}{b}\left(\frac{M_G+M_{ADM}}{2}\right),
\end{equation}
Note that for the Schwarzschild metric ($M_G=M_{ADM}$) the gravitational mass produces half of the total deflection, which corresponds to the prediction the equivalence principle.
The predictions of the equivalence principle for the deflection of light and gravitational redshift, therefore, correspond to setting the mass equal to $M_G$ and setting $M_{ADM}=0$.
  
The Shapiro time delay is given by
\begin{equation}
\Delta T=4GM\left(\frac{1+\gamma}{2}\right)\left[\ln\left(\frac{4r_Er_p}{r_0^2}\right) +1\right],
\end{equation}
where $r_E$ is the radius of the Earth's orbit, $r_p$ is the radius of the orbit of the planet used to reflect the radar signal and $r_0$ is the minimum distance of the radar beam from the Sun. In terms of $M_G$ and $M_E$ this becomes
\begin{equation}
\Delta T=4G\left(\frac{M_G+M_{ADM}}{2}\right)\left[\ln\left(\frac{4r_Er_p}{r_0^2}\right) +1\right].
\end{equation}
Thus, in a Schwarzschild spacetime the gravitational mass and the ADM mass each contribute half of the time delay.
  
Now consider the perihelion shift. In the Eddington-Roberston approach a term of order  $G^2M^2/r^2$  is added to $g_{tt}$. This term does not effect the light deflection or Shapiro time delay (to lowest order), but it does effect the perihelion shift. 
In this paper I have taken $g_{tt}\simeq -(1-2GM_G/r)$ at large distances and not considered higher order terms. It is not possible to deduce what portion of the perihelion shift is due to $M_G$ and what portion is due to $M_{ADM}$ without knowing the higher order terms. 
 
\section{Conclusion}
In this paper I examined various mass definitions in asymptotically flat spacetimes with a metric given by
\begin{equation}
ds^2\simeq-\left[1-\frac{2GM_1}{r}\right]dt^2+\left[1+\frac{2GM_2}{r}\right]dr^2+r^2d\Omega^2
\label{metcon}
\end{equation}
at large $r$, where $M_1$ and $M_2$ are constants. I showed that
\begin{equation}
M_G=M_M=M_T=M_K=M_1 \;\;\;\;\;\;\;\;\;\;\; and \;\;\;\;\;\;\;\;\;\;\;\;\; M_{\rho}=M_E=M_{LL}=M_{ADM}=M_2\;,
\end{equation}
where $M_G$ is the active gravitational mass, $M_M$ is the M$\o$ller mass, $M_T$ is the Tolamn mass, $M_K$ is the Komar mass, $M_{\rho}$ is the $\rho$ mass, $M_E$ is the Einstein mass, $M_{LL}$ is the Landau-Lifshitz mass and $M_{ADM}$ is the ADM mass. From $M_G=M_1$ it is clear that $M_1$ is the active gravitational mass of the system. 
  
The energy-momentum tensor in such spacetimes violates the weak, strong, and dominant energy conditions and, therefore, cannot be produced by an electromagnetic or a Klein-Gordon scalar field. Such spacetimes do, however, appear in string theory \cite{Gar1}, 
and in Brans-Dicke theory \cite{Brans1}.

To determine the physical significance of $M_2$ I considered a time dependent spacetime with a metric given by
\begin{equation}
ds^2\simeq-\left[1-\frac{2GM_1(t)}{r}\right]dt^2+\left[1+\frac{2 GM_2(t)}{r}\right]dr^2+r^2d\Omega^2
\label{Conc1}
\end{equation}
at large $r$ where
\begin{equation}
\lim_{t\rightarrow-\infty}M_{1,2}(t)=0
\end{equation}
and
\begin{equation}
\lim_{t\rightarrow\infty}M_{1,2}(t)=M_{1,2}.
\end{equation}
This spacetime starts as a Minkowski spacetime and evolves into the spacetime with a metric given by (\ref{Conc1}) at large $r$. The energy momentum-tensor, at large $r$, has a
$T^{t}_{\;\;\;r}$ component given by
\begin{equation}
T^{t}_{\;\;\;r}\simeq -\frac{1}{4\pi G r^2}\frac{dM_2(t)}{dt}\;.
\end{equation}
The rate at which energy flows in through a sphere of radius $R\rightarrow\infty$ is
\begin{equation}
\frac{dE}{dt}=\frac{dM_2(t)}{dt}
\end{equation}
implying that total energy that flows in through the sphere is given by
\begin{equation}
E=M_2\;.
\end{equation}
Since the energy-momentum tensor contains the inertial mass, not the gravitational mass, $M_2$ is the amount of inertial mass that flowed into the system.
   
There are two independent conservation laws:
\begin{equation}
\partial_{\nu} J^{\;\;\nu}_{\mu}=0
\label{conserv2a}
\end{equation}
and
\begin{equation}
\frac{\partial}{\partial x^{\nu}}\left[\sqrt{-g}\left(T^{\;\;\nu}_{\mu}+t^{\;\;\nu}_{\mu}\right)\right]=0
\label{conserv1a}
\end{equation}
where $ J^{\;\;\nu}_{\mu}$ is the  M$\o$ller pseudotensor and  $t^{\;\;\nu}_{\mu}$ is the Einstein pseudotensor. The Landau-Lifshitz or Weinberg pseudotensor could be used instead of the Einstein pseudotensor. Since $M_M=M_G$, I postulate that 
({\ref{conserv2a}) corresponds to the conservation of active gravitational mass. Since $M_E$ corresponds to the inertial mass that flowed into the system, I postulate that (\ref{conserv1a}) corresponds to the conservation of inertial mass. I therefore identify $M_2$ as the inertial mass of the system. Note that the inertial and active gravitational masses do not have to be the same, but both are individually conserved. This does not imply a violation of the weak equivalence principle which requires the equality of inertial and passive gravitational masses. The Einstein equations force the inertial and active gravitational masses to be the same if the matter that extends out to large $r$ satisfies the weak, strong or dominant energy condition. If the spacetime has a metric of the form (\ref{metcon}) and is vacuum at large $r$ the Einstein equations also force the inertial and active gravitational masses to be the same. 
For the remainder of the conclusion I will denote $M_1$ by $M_G$ and $M_2$ by $M_{ADM}$.   
 
The effect of $M_G$ and $M_{ADM}$ on the motion of a test particle was determined from the geodesic equation. At large $r$ the gravitational mass gives the expected Newtonian contribution, but the ADM mass only produces corrections to the Newtonian force. If $M_G=0$ a particle at rest will not experience a ``force". This is hidden in Schwarzschild spacetimes because the ADM mass has the same value as the gravitational mass.
  
I next examined the contributions of $M_G$ and $M_{ADM}$ to the gravitational redshift, the deflection of light, the Shapiro time delay and the precession of perihelia. I showed that, to lowest order, only $M_G$ contributes to the gravitational redshift while $M_G$ and $M_{ADM}$ contribute equally to the deflection of light. Thus, the predictions of the equivalence principle correspond to setting the mass equal to $M_G$ and setting $M_{ADM}=0$. I also showed that $M_G$ and $M_{ADM}$ contribute equally to the Shapiro time delay. The precession of perihelia depend on a higher order term in $g_{tt}$, so this term must be chosen before the contributions of $M_G$ and $M_{ADM}$ can be determined.
  
\section*{Acknowledgements}
This research was supported by the  Natural Sciences and Engineering Research
Council of Canada.

\end{document}